\documentclass[english]{article}
\usepackage[T1]{fontenc}
\usepackage[latin9]{inputenc}
\usepackage{amssymb}

\makeatletter
\newcommand{\lyxaddress}[1]{
\par {\raggedright #1
\vspace{1.4em}
\noindent\par}
}

\makeatother

\usepackage{babel}
\begin{document}

\title{\textbf{Third gravitational wave polarization mode in Rastall theory
and analogy with $f(R)$ theories}}

\author{\textbf{H. Moradpour$^{1}$, C. Corda$^{1}$ and Ignazio Licata$^{2,3}$}}
\maketitle

\lyxaddress{\textbf{$^{1}$}Research Institute for Astronomy and Astrophysics
of Maragha (RIAAM), Maragha 55134-441, Iran}

\lyxaddress{\begin{center}
$^{2}$ISEM, Inst. For Scientific Methodology, PA, Italy \& $^{3}$School
of Advanced International Studies on Applied Theoretical and Non Linear
Methodologies in Physics, Bari (Italy) 
\par\end{center}}
\begin{abstract}
The recent starting of the gravitational wave (GW) astronomy with
the events GW150914, GW151226, GW170104, and the very recent GW170814
and GW170817 seems to be fundamental not only in order to obtain new
intriguing astrophysical information from our surrounding Universe,
but also in order to discriminate among Einstein's general theory
of relativity (GTR) and alternative gravitational theories. At the
present time, despite the cited events, and in particular the last
ones, which are the events GW170814 and GW170817, have put very strong
constraint on the GTR, extended theories of gravity have not been
completely ruled out. Here we discuss, in our knowledge for the first
time in the literature, GWs in the Rastall theory of gravity. In fact,
the Rastall theory recently obtained a renovated interest in the literature.
We show that there is a profound analogy between GWs in $f(R)$ theories
of gravity and GWs in the Rastall theory. This will permit us to linearize
the Rastall field equations and to find the corresponding GWs. We
will also study the motion of the test masses due to GWs in this theory
which could help, in principle, to discriminate between the GTR, $f(R)$
theories and the Rastall theory of gravity.
\end{abstract}

\section{Introduction}

The observations of GWs from binary black hole (BH) mergers, i.e.
the events GW150914 \cite{key-1}, GW151226 \cite{key-2}, GW170104
\cite{key-3} and the recent GW170814 \cite{key-48}, and the very
recent observation of GWs from a neutron star (NS) merger, i.e. the
event GW170817 \cite{key-49}, represented the starting of the era
of the GW astronomy. These remarkable GW detections are unanimously
considered a cornerstone for science and for gravitational physics
in particular. On one hand, the great result is to discover new, intriguing
information on the Universe. On the other hand, the nascent GW astronomy
could be useful in order to discriminate, in an ultimate way, among
the GTR and potential alternative theories \cite{key-4}. Extended
gravity {[}4 - 7{]} is indeed an useful and popular tool to attempt
to understand the big puzzles in the standard model of cosmology like
the well known dark energy \cite{key-8,key-9} and dark matter \cite{key-10,key-11}
problems. In this framework, it is important stressing that all of
the potential alternatives to the GTR must be viable. In other words,
such alternatives must be metric theories in order to be in agreement
with the Einstein's equivalence principle, which is today supported
by a very strong empirical evidence \cite{key-5}. In addition, as
they must pass the solar system tests, deviation from the standard
GTR must be weak \cite{key-4}. 

Among various different proposals and attempting to extend the physical
framework of the GTR, the theory proposed by P. Rastall in 1972 \cite{key-12}
recently gained a renewed interest in the literature {[}13 - 16{]}.
The Rastall theory has indeed some good behavior. It shows a good
agreement with observational data on the Universe age and on the Hubble
parameter \cite{key-17}. It can, in principle, provide an alternative
description for the matter dominated era with respect to the GTR \cite{key-18}.
It is in agreement with observational data from the helium nucleosynthesis
\cite{key-19}. All these evidences have motivated physicists to study
the various cosmic eras in this framework {[}20 - 24{]}. In addition,
it seems to do not suffer from the entropy and age problems which
appear in the framework of standard cosmology \cite{key-25}. It seems
also consistent with the gravitational lensing phenomena \cite{key-26,key-27}.
Further details on the Rastall theory can be found in {[}29 - 33{]}
and references within.

A key point on the Rastall theory is that it is a theory which considers
a non-divergence-free energy-momentum {[}12 - 33{]}. \emph{The curvature-matter
theory of gravity }{[}34 - 38{]} works in a similar way. This theory
is indeed similar to the Rastall theory in a way that the matter and
geometry are coupled to each other in a non-minimal way\emph{ }{[}34
- 38{]}. Hence, the standard energy-momentum conservation law does
not work {[}34 - 38{]}.

In the following we discuss GWs in the Rastall theory also finding
the motion of the test masses due to such GWs. This could help, in
principle, to discriminate between the GTR and the Rastall theory
in the developing of the GW astronomy. It is important to stress that
despite the various GW detections, and in particular the last ones,
that are the events GW170814 \cite{key-48} and GW170817 \cite{key-49},
have put very strong constraint on the GTR, extended gravity theories
have not been completely ruled out. In fact, despite current tests
are strongly in favor of the purely GTR polarizations against scalar
polarizations which are present in extended theories \cite{key-48},
for example in $f(R)$ gravity \cite{key-51}, binary BH systems are
not at all promising for studying such scalar polarizations because
of a consequence of the no-hair theorems for BHs, see \cite{key-40,key-50}.
BHs indeed radiate away any scalar field, so that a binary BH system
in $f(R)$ gravity behaves as in the GTR. We show in this paper that
the situation is analogous in the Rastall theory, where a scalar GW
component is present. Similarly, binary NS systems, like the event
GW170817, are also not effective testing grounds for scalar radiation
\cite{key-40,key-50}. This is because NS masses tend to cluster around
the Chandrasekhar limit of $1.4M_{\circledcirc}$, being $M_{\circledcirc}$
the solar mass, and the sensitivity of NSs is not a strong function
of mass for a given equation of state \cite{key-40,key-50}. Thus,
in systems like the binary NSs, scalar radiation is naturally suppressed
by symmetry, and the bound achievable cannot compete with those from
the solar system \cite{key-40,key-50}. Hence the most promising binary
systems are mixed: BH-NS, BH-WD or NS-WD \cite{key-40,key-50}. GWs
from those mixed systems have not been yet detected by the ground
based GW interferometers.

\section{The Rastall theory of gravity}

In the Rastall theory, the ordinary energy-momentum conservation law
is modified as \cite{key-12} 
\begin{eqnarray}
T_{\ \ ;\mu}^{\mu\nu}=\lambda R^{,\nu},\label{rastal}
\end{eqnarray}
where $\lambda$ and $R$ are the Rastall constant parameter and the
Ricci scalar, respectively. Taking into account the Bianchi identity,
one gets the Rastall field equations as \cite{key-12} 
\begin{eqnarray}
G_{\mu\nu}+\kappa\lambda g_{\mu\nu}R=\kappa T_{\mu\nu},\label{r1}
\end{eqnarray}
where $\kappa$ is the Rastall gravitational coupling constant. Combining
eqs. ($\ref{r1}$) and ($\ref{rastal}$) one gets $R(4\kappa\lambda-1)=\kappa T$
and

\begin{eqnarray}
T_{\ \ ;\mu}^{\mu\nu} & = & \frac{\kappa\lambda}{4\kappa\lambda-1}T^{,\nu},\label{ein0}
\end{eqnarray}
respectively. Thus, for traceless solutions, that is $T=R=0$, the
Rastall field equations reduce to the standard GRT field equations.

Now, let us consider a congruence of geodesics of parameter $\tau$,
which is distinguished by the parameter $\Lambda$ in a way that its
tangent vector field ($v^{\alpha}$) and the separation vector ($\xi^{\alpha}$)
between the geodesics curves are evaluated as

\begin{eqnarray}
v^{\alpha}=\frac{dx^{\alpha}}{d\tau},\label{tan}
\end{eqnarray}
and 
\begin{eqnarray}
\xi^{\alpha}=\frac{dx^{\alpha}}{d\Lambda},\label{tan-1}
\end{eqnarray}
respectively \cite{key-39}. In this situation, the geodesic equation
reads \cite{key-39,key-40}

\begin{eqnarray}
\frac{D^{2}\xi^{\alpha}}{D\tau^{2}}=R_{\beta\mu\nu}^{\alpha}v^{\beta}v^{\mu}\xi^{\nu},\label{gde1}
\end{eqnarray}
where $R_{\beta\mu\nu}^{\alpha}$ is the Reimann tensor and \cite{key-39,key-40}
\begin{eqnarray}
\frac{D^{2}\xi^{\alpha}}{D\tau^{2}}\equiv\frac{d^{2}\xi^{\alpha}}{d\tau^{2}}+\Gamma_{\mu\nu}^{\alpha}\frac{d\xi^{\mu}}{d\tau}\frac{d\xi^{\nu}}{d\tau}.\label{eq: GDE}
\end{eqnarray}
Eq. ($\ref{gde1}$) shows that the the geodesics are bent by the space-time
curvature. The Reimann tensor can be expanded as 
\begin{equation}
\begin{array}{c}
R_{\alpha\beta\gamma\delta}=\frac{\bigl(g_{\alpha\gamma}R_{\delta\beta}-g_{\alpha\delta}R_{\gamma\beta}+g_{\beta\delta}R_{\gamma\alpha}-g_{\beta\gamma}R_{\delta\alpha}\bigr)}{2}\\
\\
-\frac{R}{6}\bigl(g_{\alpha\gamma}g_{\delta\beta}-g_{\alpha\delta}g_{\gamma\beta}\bigr)+C_{\alpha\beta\gamma\delta},
\end{array}\label{ReimannC0}
\end{equation}
where $C_{\alpha\beta\gamma\delta}$ is the Weyl tensor \cite{key-39}.
In addition, if one uses Eq. ($\ref{r1}$), one gets 

\begin{eqnarray}
R_{\alpha\beta} & = & \kappa[E_{\alpha\beta}+\frac{\kappa\lambda}{4\kappa\lambda-1}Tg_{\alpha\beta}],\label{RiemannC}
\end{eqnarray}
where 
\begin{equation}
E_{\alpha\beta}=T_{\alpha\beta}-\frac{1}{2}Tg_{\alpha\beta}\label{eq: Einstein part}
\end{equation}
is the Einsteinian part. This equation covers the Einsteinian results
at the $\lambda\rightarrow0$ limit. Finally, for a space-time of
metric $g_{\alpha\beta}$ filled by an energy-momentum source $T_{\alpha\beta}$,
Eq. ($\ref{ReimannC0}$) can be rewritten as

\begin{eqnarray}
R_{\alpha\beta\gamma\delta}=\kappa\mathcal{R}_{\alpha\beta\gamma\delta}+C_{\alpha\beta\gamma\delta}+\kappa\lambda\tilde{R}_{\alpha\beta\gamma\delta}\label{RiemannC2}
\end{eqnarray}
where

\begin{eqnarray}
\mathcal{R}_{\alpha\beta\gamma\delta} & = & \frac{\bigl(g_{\alpha\gamma}E_{\delta\beta}-g_{\alpha\delta}E_{\gamma\beta}+g_{\beta\delta}E_{\gamma\alpha}-g_{\beta\gamma}E_{\delta\alpha}\bigr)}{2}\nonumber \\
 & + & \frac{T}{6}\bigl(g_{\alpha\gamma}g_{\delta\beta}-g_{\alpha\delta}g_{\gamma\beta}\bigr),
\end{eqnarray}
and $T=\frac{4\kappa\lambda-1}{\kappa}R$. Moreover,

\begin{eqnarray}
\kappa\lambda\tilde{R}_{\alpha\beta\gamma\delta} & = & \kappa\lambda[\frac{\tilde{T}}{6}\bigl(g_{\alpha\gamma}g_{\delta\beta}-g_{\alpha\delta}g_{\gamma\beta}\bigr)\label{RiemannC3}\\
 & + & \frac{\bigl(g_{\alpha\gamma}\tilde{E}_{\delta\beta}-g_{\alpha\delta}\tilde{E}_{\gamma\beta}+g_{\beta\delta}\tilde{E}_{\gamma\alpha}-g_{\beta\gamma}\tilde{E}_{\delta\alpha}\bigr)}{2}]\nonumber 
\end{eqnarray}
is the correction term which directly comes from the Rastall hypothesis.
In the above equations 
\begin{eqnarray}
\tilde{E}_{\alpha\beta} & = & \frac{\kappa}{4\kappa\lambda-1}Tg_{\alpha\beta},\nonumber \\
\tilde{T} & = & \tilde{E}_{\alpha}^{\alpha}=\frac{4\kappa}{4\kappa\lambda-1}T=4R.
\end{eqnarray}
Hence, for the traceless sources, such as the radiation field, where
$T=0$, we have 
\begin{equation}
\tilde{R}_{\alpha\beta\gamma\delta}=0.
\end{equation}
In the Rastall theory, $\kappa$ differs from the Einstein gravitational
coupling ($\kappa_{E}$) \cite{key-12,key-17,key-18}. Thus, we have
\begin{equation}
\kappa=\frac{4\gamma-1}{6\gamma-1}\kappa_{E},
\end{equation}
where $\gamma\equiv\kappa\lambda$ is the dimensionless Rastall parameter
\cite{key-13}. Inserting all of the above results in Eq. ($\ref{gde1}$),
one gets 
\begin{eqnarray}
\frac{D^{2}\xi^{\alpha}}{D\tau^{2}}=(\frac{D^{2}\xi^{\alpha}}{D\tau^{2}})_{E}+(\frac{D^{2}\xi^{\alpha}}{D\tau^{2}})_{R},\label{gde2}
\end{eqnarray}
where

\begin{eqnarray}
(\frac{D^{2}\xi^{\alpha}}{D\tau^{2}})_{E} & = & [\kappa_{E}\mathcal{R}_{\beta\mu\nu}^{\alpha}+C_{\beta\mu\nu}^{\alpha}]v^{\beta}v^{\mu}\xi^{\nu},\label{gde3}
\end{eqnarray}
and

\begin{eqnarray}
(\frac{D^{2}\xi^{\alpha}}{D\tau^{2}})_{R} & = & \gamma[\tilde{R}_{\beta\mu\nu}^{\alpha}+\frac{2}{1-6\gamma}\mathcal{R}_{\beta\mu\nu}^{\alpha}]v^{\beta}v^{\mu}\xi^{\nu},\label{RiemannC4}
\end{eqnarray}
are the geodesic equations in the Einstein framework and the correction
term to the geodesic equation due to the Rastall hypothesis, respectively.
Clearly, the results of the GTR are obtainable in the appropriate
limit of $\lambda\rightarrow0$ parallel to the $\gamma=0$ limit.
From now, we work in units setting $\kappa_{E}=1.$ Thus, $\kappa=\frac{4\gamma-1}{6\gamma-1}.$ 

\section{Linearized Rastall field equations}

In this Section, GWs in the Rastall framework will be analysed. For
the sake of simplicity, we work with $G=1$, $c=1$ and $\hbar=1$
(natural units) in the following. We consider a weak field situation
where the observer is too far away from the energy-momentum source.
In other words, we investigate a space-time in which the metric $g_{\mu\nu}$
can be expanded as \cite{key-41}

\begin{eqnarray}
g_{\mu\nu}=\eta_{\mu\nu}+h_{\mu\nu},\label{met}
\end{eqnarray}
where $\eta_{\mu\nu}$ is the Minkowski metric and $h_{\mu\nu}$ denote
the deviation from the Minkowski space-time due to the curvature carried
by the GW. Applying Eq. ($\ref{met}$) to Eq. ($\ref{r1}$) and following
the approach of  \cite{key-41,key-42}, one finds 

\begin{eqnarray}
 &  & h_{\ \nu\ ,\alpha\mu}^{\alpha}+h_{\ \mu\ ,\alpha\nu}^{\alpha}-h_{,\mu\nu}-\Box h_{\mu\nu}\nonumber \\
 &  & -(1-2\gamma)\eta_{\mu\nu}(h_{\ \ ,\alpha\beta}^{\alpha\beta}-\Box h)=2\kappa T_{\mu\nu}.\label{eq: 21}
\end{eqnarray}
Let us simplify Eq. (\ref{eq: 21}), in similarity to the case of
the GTR. One defines \cite{key-41,key-42}

\begin{eqnarray}
\bar{h}_{\mu\nu}=h_{\mu\nu}-\frac{1}{2}\eta_{\mu\nu}h.\label{gw3}
\end{eqnarray}
Combining Eq. (\ref{gw3}) with Eq. ($\ref{eq: 21}$) one gets

\begin{eqnarray}
 &  & 2\kappa T_{\mu\nu}=\bar{h}_{\ \nu,\alpha\mu}^{\alpha}+\bar{h}_{\ \mu,\alpha\nu}^{\alpha}-\Box\bar{h}_{\mu\nu}\label{gw4}\\
 &  & +\gamma\eta_{\mu\nu}\Box\bar{h}+(1-2\gamma)\eta_{\mu\nu}\bar{h}_{\ \ ,\alpha\beta}^{\alpha\beta}\ ,\nonumber 
\end{eqnarray}
which leads to

\begin{eqnarray}
 &  & 2\kappa T_{\mu\nu}=\gamma\eta_{\mu\nu}\Box\bar{h}-\Box\bar{h}_{\mu\nu}\label{gw5}
\end{eqnarray}
Considering the Lorenz condition \cite{key-43} $\bar{h}_{\ \nu,\alpha}^{\alpha}=0$,
for either a vacuum spacetime or a long distance from the source we
have

\begin{eqnarray}
 &  & \Box\bar{h}_{\mu\nu}=\gamma\eta_{\mu\nu}\Box\bar{h}.\label{gw7}
\end{eqnarray}
One can also check that, at the $\gamma\rightarrow0$ limit, the Einstein
linearized field equations are re-obtained \cite{key-41,key-42}.

Now, let us define 

\begin{eqnarray}
\bar{h}_{\mu\nu}-\gamma\eta_{\mu\nu}\bar{h}\equiv\bar{H}_{\mu\nu}.\label{gws1}
\end{eqnarray}
Combining Eq. (\ref{gws1}) with Eq. ($\ref{gw3}$\}) one gets 

\begin{eqnarray}
h_{\mu\nu}=\bar{H}_{\mu\nu}+\frac{(1-2\gamma)\bar{H}}{2(4\gamma-1)}\eta_{\mu\nu},\label{gws2}
\end{eqnarray}
where $\bar{H}$ is the trace of $\bar{H}_{\mu\nu}$. Inserting Eq.
($\ref{gws1}$) into Eq. ($\ref{gw7}$), one easily finds

\begin{eqnarray}
 &  & \Box\bar{H}_{\mu\nu}=0.\label{gws3}
\end{eqnarray}
It is also interesting to mention here that, based on Eqs. ($\ref{gws2}$),
($\ref{gws1}$) and ($\ref{gw3}$), it seems that the traceless solutions
of Eq. ($\ref{gws3}$) are not affected by the mutual non-minimal
coupling between geometry and matter fields (or equally by the $\gamma$
parameter). This is due to the fact that the $\gamma$ parameter only
appears with the trace of solutions and thus perturbed metric. Therefore,
it seems that the transverse and traceless solutions of this equations
are exactly the same as those of the GTR case \cite{key-41,key-42}.

In the absence of energy-momentum source, the linearized approximation
of Eq. ($\ref{rastal}$) leads to

\begin{eqnarray}
h_{\ \ ,\alpha\beta}^{\alpha\beta}-\Box h=C,\label{l1}
\end{eqnarray}
where $C$ is an integration constant. On the other hand, the trace
of Eq. ($\ref{r1}$) implies $C=0$ in the absence of energy-momentum
source. Bearing the Lorenz gauge in mind \cite{key-43} and using
Eq. ($\ref{gws2}$), one immediately finds

\begin{eqnarray}
 &  & \Box\bar{H}=0,\label{l2}
\end{eqnarray}
in full agreement with Eq. ($\ref{gws3}$). One may also use the Bianchi
identity as well as Eqs. ($\ref{rastal}$) and ($\ref{gw3}$) to obtain

\begin{eqnarray}
(\frac{1}{2}[h_{,\mu\nu}-h_{,\nu\mu}]-\Box h_{\mu\nu})^{;\nu}=0,\label{l3}
\end{eqnarray}
which is compatible with Eqs. ($\ref{gws3}$) and ($\ref{l2}$). In
fact, since $h_{,\mu\nu}=h_{,\nu\mu}$, one can use Eq. ($\ref{gws2}$)
to get

\begin{eqnarray}
\Box h_{\mu\nu}=\Box\bar{H}_{\mu\nu}+\frac{(1-2\gamma)}{2(4\gamma-1)}\eta_{\mu\nu}\Box\bar{H}=0.\label{l4}
\end{eqnarray}
This indicates that Eq. ($\ref{l3}$) does not give us anything more
than Eqs. ($\ref{gws3}$) and ($\ref{l2}$). 

\section{Gravitational waves }

It is straightforward to check that the plane waves

\begin{eqnarray}
\bar{H}_{\mu\nu}=Q_{\mu\nu}\exp(ik_{\alpha}x^{\alpha})+c.c.,\label{s1}
\end{eqnarray}
where $k_{\alpha}k^{\alpha}=0,$ are solutions of Eq. ($\ref{gws3}$).
Then, for the null wave-vector $k_{\alpha}$, one gets 

\begin{eqnarray}
k^{\alpha}=(\omega,k^{1},k^{2},k^{3}).\label{nwave}
\end{eqnarray}
Here, $\omega^{2}\equiv k_{i}k^{i}$ is the wave frequency observed
by an observer with four-velocity $U^{\alpha},$ i.e. $\omega=-k_{\alpha}U^{\alpha}$.

Now, combining Eqs. ($\ref{s1}$), ($\ref{gws1}$) and ($\ref{gws2}$),
one finds

\begin{eqnarray}
\bar{h}_{\mu\nu}=A_{\mu\nu}\exp(ik_{\alpha}x^{\alpha})+c.c.,\label{s10}
\end{eqnarray}
and

\begin{eqnarray}
h_{\mu\nu}=a_{\mu\nu}\exp(ik_{\alpha}x^{\alpha})+c.c.,\label{s20}
\end{eqnarray}
where $A_{\mu\nu}=Q_{\mu\nu}+\frac{\gamma Q}{4\gamma-1}\eta_{\mu\nu}$
and $a_{\mu\nu}=Q_{\mu\nu}+\frac{(1-2\gamma)Q}{2(4\gamma-1)}\eta_{\mu\nu}$,
respectively. Since $h_{\mu\nu}$ is symmetric, $Q_{\mu\nu}$, $A_{\mu\nu}$
and $a_{\mu\nu}$ have $10$ independent components. In addition,
bearing the Lorenz gauge ($h_{\alpha\beta}^{\ \ \ ,\alpha}=k^{\alpha}A_{\alpha\beta}=0$)
in mind \cite{key-43}, we get

\begin{eqnarray}
k^{\mu}Q_{\mu\nu}=\frac{\gamma Q}{1-4\gamma}k_{\nu}.\label{s2}
\end{eqnarray}
In the above equations, $Q(\equiv Q_{\mu}^{\mu})$ is the trace of
$Q_{\mu\nu}$. Since $k_{\alpha}$ is a null vector, Eq. ($\ref{s2}$)
leads to

\begin{eqnarray}
 &  & k^{\nu}k^{\mu}Q_{\mu\nu}=0.\label{s3}
\end{eqnarray}
It is interesting to note here that this result can also be obtained
if one uses Eq. ($\ref{l1}$). In fact, as we considered an empty
spacetime, we have $R\simeq h_{\ \ ,\mu\nu}^{\mu\nu}-\Box h=0$ at
the weak field limit \cite{key-41,key-42} which leads to $k_{\mu}k_{\nu}a^{\mu\nu}=0$
and so the above result. The Lorenz condition ($k_{\mu}A^{\mu\nu}=0$)
\cite{key-43} also reduces the independent components of $A_{\mu\nu}$
to $6$ components, and therefore, $Q_{\mu\nu}$ and $a_{\mu\nu}$
will also have $6$ independent components, a result in agreement
with Eqs. ($\ref{s2}$) and ($\ref{s3}$). This implies that the Rastall
theory should have 4 additional GW polarizations with respect to the
two standard polarizations of the GTR case \cite{key-41,key-42}.

Now, inserting $\omega=-k_{\alpha}U^{\alpha}$ into Eq. ($\ref{s2}$),
it is also easy to obtain

\begin{eqnarray}
 &  & U^{\nu}k^{\mu}Q_{\mu\nu}=\frac{Q\gamma\omega}{4\gamma-1},\label{s4}
\end{eqnarray}
which recovers the GTR result in the appropriate limit of $\gamma=0$.
Theoretically, this equation should also reduces the number of independent
components of $Q_{\mu\nu}$ to $2$. But, as $U^{\alpha}$ and the
wave-vector field are in the mutual relation $\omega=-k_{\alpha}U^{\alpha}$,
one component of the $U^{\alpha}$ vector field is not arbitrary and
free. Thus, Eq. ($\ref{s4}$) may reduce independent components of
$Q_{\mu\nu}$, $A_{\mu\nu}$ and $a_{\mu\nu}$ to maximum $3$ components.
In fact, Eqs. ($\ref{s3}$) and ($\ref{s4}$) specify relations between
the wave amplitude, the wave-vector and the four-velocity of the observer.
Thus, the Rastall theory has one additional GW polarization with respect
to the 2 standard polarizations of the GTR case \cite{key-41,key-42}.
We notice that there is a strong analogy with GWs in $f(R)$ theories
of gravity. In fact, in that case the linearized field equations are
\cite{key-51}

\begin{equation}
\begin{array}{c}
\square\bar{h}_{\mu\nu}=0\\
\\
\square h_{f}=m^{2}h_{f}
\end{array}\label{eq: onde f(R)}
\end{equation}
where $h_{f}$ is a massive effective scalar field representing the
third GW polarization in $f(R)$ theories of gravity. $h_{f}$ is
due to the presence of curvature high order terms in the $f(R)$ action
and is obtained by taking the trace of the field equations. In the
current case of the Rastall theory, if one gathers Eq. ($\ref{gws3}$)
with Eq. (\ref{l2}) one gets 
\begin{equation}
\begin{array}{c}
\Box\bar{H}_{\mu\nu}=0\\
\\
\Box\bar{H}=0
\end{array}\label{eq: onde Rastall}
\end{equation}
and we can interpret $\bar{H}$ as being a massless effective scalar
field representing the third GW polarization in the Rastall theory.
In fact, also in Rastall theory $\bar{H}$ is obtained by taking the
trace of the field equations and its physical origin arises from the
curvature term in Eq. (\ref{r1}). Thus, in order to further go ahead
in the linearization process, we can follow the analysis in \cite{key-51},
but keeping in mind that now the effective scalar field is massless
rather than massive. 

The solutions of the first of Eqs. (\ref{eq: onde Rastall}) are the
plane waves (\ref{s1}). One must add the solutions of the second
of Eqs. (\ref{eq: onde Rastall}) obtaining 
\begin{equation}
\bar{H}_{\mu\nu}=Q_{\mu\nu}(\overrightarrow{k})\exp(ik^{\alpha}x_{\alpha})+c.c.\label{eq: sol T}
\end{equation}

\begin{equation}
\bar{H}=Q(\overrightarrow{k})\exp(ik^{\alpha}x_{\alpha})+c.c.\label{eq: sol S}
\end{equation}
The solutions (\ref{eq: sol T}) and (\ref{eq: sol S}) take the conditions
(\ref{s2}). One considers a GW propagating in the positive $z$ direction
with

\begin{equation}
k^{\mu}=(k,0,0k).\label{eq: k}
\end{equation}
Eqs. (\ref{s2}) imply

\begin{equation}
\begin{array}{c}
Q_{0\nu}=-Q_{3\nu}\\
\\
Q_{\nu0}=-Q_{\nu3}\\
\\
Q_{00}=-Q_{30}+Q_{33}.
\end{array}\label{eq: A}
\end{equation}
One recalls that the freedom degrees of $Q_{\mu\nu}$ are 3. In fact,
we started with 10 components ($Q_{\mu\nu}$ is a symmetric tensor).
As it has been stressed above, the Lorenz condition $k_{\mu}A^{\mu\nu}=0$
\cite{key-43} reduces the components to 6. Then, we take $Q_{00}$,
$Q_{11}$, $Q_{22}$, $Q_{21}$, $Q_{31}$, $Q_{32}$ like independent
components. The condition ($\ref{s4}$) sets to zero 3 more components.
Now, one takes 

\begin{equation}
\begin{array}{c}
\epsilon_{\mu}=\tilde{\epsilon}_{\mu}(\overrightarrow{k})\exp(ik^{\alpha}x_{\alpha})+c.c.\\
\\
k^{\mu}\tilde{\epsilon}_{\mu}=0,
\end{array}\label{eq: ancora gauge}
\end{equation}
with the condition $\Box\epsilon_{\nu}=\partial^{\mu}\bar{H}_{\mu\nu}$
for the parameter $\epsilon^{\mu}$. Then, the transform law for $Q_{\mu\nu}$
reads (one considers again the Lorenz condition \cite{key-43} and
Eq. (\ref{eq: sol T}) )

\begin{equation}
Q_{\mu\nu}\rightarrow Q'_{\mu\nu}=Q_{\mu\nu}-2ik(_{\mu}\tilde{\epsilon}_{\nu}).\label{eq: trasf. tens.}
\end{equation}
Hence, one can write down the six components of interest as

\begin{equation}
\begin{array}{ccc}
Q_{00} & \rightarrow & Q_{00}+2ik\tilde{\epsilon}_{0}\\
Q_{11} & \rightarrow & Q_{11}\\
Q_{22} & \rightarrow & Q_{22}\\
Q_{21} & \rightarrow & Q_{21}\\
Q_{31} & \rightarrow & Q_{31}-ik\tilde{\epsilon}_{1}\\
Q_{32} & \rightarrow & Q_{32}-ik\tilde{\epsilon}_{2}.
\end{array}\label{eq: sei tensori}
\end{equation}
Clearly, the components of $Q_{\mu\nu}$ having physical meaning are
the gauge-invariants $Q_{11}$, $Q_{22}$ and $Q_{21}$. Thus, one
chooses $\tilde{\epsilon}_{\nu}$ to set equal to zero the others.
The massless effective scalar field is obtained as

\begin{equation}
\bar{H}=\bar{H}_{11}+\bar{H}_{22}.\label{eq: trovato scalare}
\end{equation}
Now, defining $h_{Rast}\equiv-\bar{H}$, the total GW perturbation
propagating in the $z+$ direction reads

\begin{equation}
h_{\mu\nu}(t-z)=Q^{+}(t-z)e_{\mu\nu}^{(+)}+Q^{\times}(t-z)e_{\mu\nu}^{(\times)}+h_{Rast}(t-z)e_{\mu\nu}^{(Rast)}.\label{eq: perturbazione totale}
\end{equation}

The term $Q_{+}(t-z)e_{\mu\nu}^{(+)}+Q_{\times}(t-z)e_{\mu\nu}^{(\times)}$
represents the two standard GW polarizations of the GTR in the transverse-traceless
gauge \cite{key-41} while the term $h_{Rast}(t-z)e_{\mu\nu}^{(Rast)}$
is the third additional polarization due to the curvature term in
Eq. (\ref{r1}). 

\section{Geodesic deviation equation and test masses motion}

From Eq. (\ref{eq: perturbazione totale}) one gets the total GW line-element
as 
\begin{equation}
ds^{2}=dt^{2}-dz^{2}-(1+Q_{+}+h_{Rast})dx^{2}-(1-Q_{+}+h_{Rast})dy^{2}-2Q_{\times}dxdy.\label{eq: metrica TT super totale}
\end{equation}
As the GW astronomy is performed in a laboratory environment on Earth,
one usually uses the coordinate system in which the space-time is
locally flat \cite{key-41}. In that case, the distance between any
two points and/or test masses is given simply by the difference in
their coordinates in the sense of Newtonian physics \cite{key-41}.
In this gauge, called the gauge of the local observer, GWs manifest
themselves by exerting tidal forces on the test masses, which are
the mirror and the beam-splitter in the case of an interferometer
like LIGO. A complete analysis of gauge of the local observer is given
in \cite{key-41}. Here we limit ourselves to recall only the more
important behaviors of this gauge:
\begin{enumerate}
\item The proper time of the observer O is given by the time coordinate
$x_{0}$.
\item Spatial axes are centered in O.
\item If both of acceleration and rotation are null, then the spatial coordinates
$x_{j}$ are the proper distances along the axes. In that case the
gauge of the local observer reduces to a local Lorentz gauge and the
metric is 
\begin{equation}
ds^{2}=(-dx^{0})^{2}+\delta_{ij}dx^{i}dx^{j}+O(|dx^{j}|^{2})dx^{\alpha}dx^{\beta};\label{eq: metrica local lorentz}
\end{equation}
\item The effect of GWs on test masses is described by the geodesic deviation
equation 
\begin{equation}
\ddot{x^{i}}=-\widetilde{R}_{0k0}^{i}x^{k},\label{eq: deviazione geodetiche}
\end{equation}
being $\widetilde{R}_{0k0}^{i}$ the linearized Riemann tensor \cite{key-41}.
\end{enumerate}
The effect on test masses due to the two standard GTR polarizations
$Q_{+}$ and $Q_{\times}$is well known \cite{key-41}. Thus, we will
consider only the effect on test masses by the additional polarization
$h_{Rast}.$ In that case, the line element (\ref{eq: metrica TT super totale})
reduces to 
\begin{equation}
ds^{2}=dt^{2}-dz^{2}-(1+h_{Rast})dx^{2}-(1+h_{Rast})dy^{2}.\label{eq: metrica Rastall}
\end{equation}
In order to further stress the analogy with GWs in $f(R)$ theories
of gravity, one makes the following coordinate transformation 
\begin{equation}
\begin{array}{ccc}
dx' & = & dx\\
dy' & = & dy\\
dz' & = & (1+\frac{1}{2}h_{Rast})dz-\frac{1}{2}h_{Rast}dt\\
cdt' & = & (1+\frac{1}{2}h_{Rast})dt-\frac{1}{2}h_{Rast}dz.
\end{array}\label{eq: transf coordinate}
\end{equation}
Then, the new line element is the conformally flat one 
\begin{equation}
ds^{2}=\left[1+h_{Rast}(t-z)\right](-dt^{2}+dz^{2}+dx^{2}+dy^{2}).\label{eq: metrica conformemente piatta}
\end{equation}
The new gauge of Eq. (\ref{eq: metrica conformemente piatta}) is
very similar to the one which is usually used to discuss GWs in $f(R)$
theories of gravity, that is \cite{key-51,key-53} 
\begin{equation}
ds^{2}=\left[1+h_{f}(t-v_{G}z)\right](-dt^{2}+dz^{2}+dx^{2}+dy^{2}).\label{eq: metrica puramente scalare}
\end{equation}
In fact, we recall that the third GW polarization in $f(R)$ theories
of gravity is interpreted in terms of a massive field which can be
discussed like a wave-packet \cite{key-51,key-53} . The group-velocity
of a wave-packet of $h_{f}$ centered in $\overrightarrow{p}$ is 

\begin{equation}
\overrightarrow{v_{G}}=\frac{\overrightarrow{p}}{\omega},\label{eq: velocita' di gruppo}
\end{equation}
which is exactly the velocity of a massive particle with mass $m$
and momentum $\overrightarrow{p}$. The group-velocity results \cite{key-51,key-53} 

\begin{equation}
v_{G}=\frac{\sqrt{\omega^{2}-m^{2}}}{\omega},\label{eq: velocita' di gruppo 2}
\end{equation}
and, if one wants a constant speed of the wave-packet, one gets \cite{key-51,key-53} 

\begin{equation}
m=\sqrt{(1-v_{G}^{2})}\omega.\label{eq: relazione massa-frequenza}
\end{equation}
Thus, from Eq. (\ref{eq: metrica conformemente piatta}), one sees
that the speed of the third additional massless GW mode in the Rastall
theory is exactly the speed of light, while, from Eq. (\ref{eq: metrica puramente scalare})
one sees that the speed of the third additional massive GW mode in
$f(R)$ theories of gravity is the group-velocity $v_{G}<c,$ as one
expects. 

Now, we want to study the test mass motion in the presence of the
third GW polarization in the Rastall theory of gravity. In order to
achieve this, one could derive the coordinates transformation from
the line element (\ref{eq: metrica conformemente piatta}) to the
local Lorentz line element (\ref{eq: metrica local lorentz}). This
is, in principle, possible, because an analogous coordinates transformation
is well known in the standard GTR case, see for example \cite{key-52}.
On the other hand, this is not necessary, because the linearized Riemann
tensor is invariant under gauge transformations \cite{key-41,key-53}.
Hence, it can be directly computed from Eq. (\ref{eq: metrica conformemente piatta}).
Again, we can use the analogy with GWs in $f(R)$ theories of gravity
and perform a computation similar to the one in \cite{key-53}, but
keeping in mind the the field is now massless rather than massive.
From \cite{key-41,key-53} one gets 
\begin{equation}
\widetilde{R}_{\mu\nu\rho\sigma}=\frac{1}{2}\{\partial_{\mu}\partial_{\beta}h_{\alpha\nu}+\partial_{\nu}\partial_{\alpha}h_{\mu\beta}-\partial_{\alpha}\partial_{\beta}h_{\mu\nu}-\partial_{\mu}\partial_{\nu}h_{\alpha\beta}\},\label{eq: riemann lineare}
\end{equation}
that, in the case eq. (\ref{eq: metrica conformemente piatta}), reads 

\begin{equation}
\widetilde{R}_{0\gamma0}^{\alpha}=\frac{1}{2}\{\partial^{\alpha}\partial_{0}h_{Rast}\eta_{0\gamma}+\partial_{0}\partial_{\gamma}h_{Rast}\delta_{0}^{\alpha}-\partial^{\alpha}\partial_{\gamma}h_{Rast}\eta_{00}-\partial_{0}\partial_{0}h_{Rast}\delta_{\gamma}^{\alpha}\}.\label{eq: riemann lin scalare}
\end{equation}
One writes the different elements as (only the non zero ones will
be considered)

\begin{equation}
\partial^{\alpha}\partial_{0}h_{Rast}\eta_{0\gamma}=\left\{ \begin{array}{ccc}
\partial_{t}^{2}h_{Rast} & for & \alpha=\gamma=0\\
\\
-\partial_{z}\partial_{t}h_{Rast} & for & \alpha=3;\gamma=0
\end{array}\right\} \label{eq: calcoli}
\end{equation}

\begin{equation}
\partial_{0}\partial_{\gamma}h_{Rast}\delta_{0}^{\alpha}=\left\{ \begin{array}{ccc}
\partial_{t}^{2}h_{Rast} & for & \alpha=\gamma=0\\
\\
\partial_{t}\partial_{z}h_{Rast} & for & \alpha=0;\gamma=3
\end{array}\right\} \label{eq: calcoli2}
\end{equation}

\begin{equation}
-\partial^{\alpha}\partial_{\gamma}h_{Rast}\eta_{00}=\partial^{\alpha}\partial_{\gamma}\Phi=\left\{ \begin{array}{ccc}
-\partial_{t}^{2}h_{Rast} & for & \alpha=\gamma=0\\
\\
\partial_{z}^{2}h_{Rast} & for & \alpha=\gamma=3\\
\\
-\partial_{t}\partial_{z}h_{Rast} & for & \alpha=0;\gamma=3\\
\\
\partial_{z}\partial_{t}h_{Rast} & for & \alpha=3;\gamma=0
\end{array}\right\} \label{eq: calcoli3}
\end{equation}

\begin{equation}
-\partial_{0}\partial_{0}h_{Rast}\delta_{\gamma}^{\alpha}=\begin{array}{ccc}
-\partial_{z}^{2}h_{Rast} & for & \alpha=\gamma\end{array}.\label{eq: calcoli4}
\end{equation}
Now, putting these results in eq. (\ref{eq: riemann lin scalare})
one obtains

\begin{equation}
\begin{array}{c}
\widetilde{R}_{010}^{1}=-\frac{1}{2}\ddot{h}_{Rast}\\
\\
\widetilde{R}_{010}^{2}=-\frac{1}{2}\ddot{h}_{Rast}\\
\\
\widetilde{R}_{030}^{3}=0.
\end{array}\label{eq: componenti riemann}
\end{equation}
We recall that in the case of $f(R)$ theories of gravity one gets
\cite{key-53} 
\begin{equation}
\begin{array}{c}
\widetilde{R}_{010}^{1}=-\frac{1}{2}\ddot{h}_{f}\\
\\
\widetilde{R}_{010}^{2}=-\frac{1}{2}\ddot{h}_{f}\\
\\
\widetilde{R}_{030}^{3}=\frac{1}{2}m^{2}h_{f},
\end{array}\label{eq: componenti riemann f(R)}
\end{equation}
instead. In fact, in the case of $f(R)$ theories of gravity the field
is massive and not transverse, see \cite{key-51,key-53} for details. 

Using Eqs. (\ref{eq: componenti riemann}) and (\ref{eq: deviazione geodetiche})
one gets 

\begin{equation}
\ddot{x}=\frac{1}{2}\ddot{h}_{Rast}x,\label{eq: accelerazione mareale lungo x}
\end{equation}

\begin{equation}
\ddot{y}=\frac{1}{2}\ddot{h}_{Rast}y\label{eq: accelerazione mareale lungo y}
\end{equation}
which means that in the current Rastall case the field is massless
and transverse. Thus, we consider a test mass which is free to move
in the plane $z=0$. Equations (\ref{eq: accelerazione mareale lungo x})
and (\ref{eq: accelerazione mareale lungo y}) give the tidal acceleration
of our test mass caused by the third polarization of the Rastall GW
in the $x$ direction and in the $y$ direction respectively. Following
\cite{key-41} one can equivalently say that there is a gravitational
potential given by 

\begin{equation}
V(\overrightarrow{r},t)=-\frac{1}{4}\ddot{h}_{Rast}t[x^{2}+y^{2}].\label{eq:potenziale in gauge Lorentziana}
\end{equation}
generating the tidal forces. Therefore, the motion of the test mass
is governed by the Newtonian equation

\begin{equation}
\ddot{\overrightarrow{r}}=-\bigtriangledown V.\label{eq: Newtoniana}
\end{equation}
One finds the solution of Eqs. (\ref{eq: accelerazione mareale lungo x})
and (\ref{eq: accelerazione mareale lungo y}) through the perturbation
method \cite{key-41}. To first order in the amplitude $h_{Rast}$
the displacements of the test mass due to the third polarization of
the Rastall GW are given by

\begin{equation}
\delta x(t)=\frac{1}{2}x_{0}h_{Rast}(t)\label{eq: spostamento lungo x}
\end{equation}
and

\begin{equation}
\delta y(t)=\frac{1}{2}y_{0}h_{Rast}(t),\label{eq: spostamento lungo y}
\end{equation}
where $x_{0}$ and $y_{0}$ are the initial coordinates of the test
mass, i.e. the coordinates of the test mass before the arrival of
the Rastall GW. If one considers again the analogy with GWs in $f(R)$
theories of gravity, one recalls that the displacements of the test
mass due to the third polarization of the $f(R)$ GW are instead given
by \cite{key-53} 
\begin{equation}
\begin{array}{c}
\delta x(t)=\frac{1}{2}x_{0}h_{f}(t)\\
\\
\delta y(t)=\frac{1}{2}y_{0}h_{f}(t)\\
\\
\delta z(t)=\frac{(1-v_{G}^{2})}{2}z_{0}h_{f}(t).
\end{array}\label{eq: spostamento f(R)}
\end{equation}
Thus, one see that, differently from the current case of the Rastall
theory, in $f(R)$ theories of gravity a longitudinal component is
present. In fact, in that case, the effect of the mass is the generation
of a \textit{longitudinal} force in addition to the transverse one,
see \cite{key-53}. 

Hence, one sees that the total displacements of the test mass due
to a GW in Rastall theory are different with respect to the total
displacements of the test mass due to a GW in the standard GTR and
with respect to the total displacements of the test mass due to a
GW in $f(R)$ theories of gravity. In the case of the GTR one indeed
obtains \cite{key-41} 
\begin{equation}
\begin{array}{c}
\delta x(t)=\frac{1}{2}[x_{0}Q_{+}(t)-y_{0}Q_{\times}(t)]\\
\\
\delta y(t)=-\frac{1}{2}[y_{0}Q_{+}(t)+x_{0}Q_{\times}(t)]\\
\\
\delta z(t)=0.
\end{array}\label{eq: traditional GTR}
\end{equation}
In the case of $f(R)$ theories of gravity one gets \cite{key-53}
\begin{equation}
\begin{array}{c}
\delta x(t)=\frac{1}{2}[x_{0}Q_{+}(t)-y_{0}Q_{\times}(t)]+\frac{1}{2}x_{0}h_{f}(t)\\
\\
\delta y(t)=-\frac{1}{2}[y_{0}Q_{+}(t)+x_{0}Q_{\times}(t)]+\frac{1}{2}y_{0}h_{f}(t)\\
\\
\delta z(t)=\frac{(1-v_{G}^{2})}{2}z_{0}h_{f}(t).
\end{array}\label{eq: f(R) case}
\end{equation}
Finally, in the case of the Rastall theory of gravity one obtains
\begin{equation}
\begin{array}{c}
\delta x(t)=\frac{1}{2}[x_{0}Q_{+}(t)-y_{0}Q_{\times}(t)]+\frac{1}{2}x_{0}h_{Rast}(t)\\
\\
\delta y(t)=-\frac{1}{2}[y_{0}Q_{+}(t)+x_{0}Q_{\times}(t)]+\frac{1}{2}y_{0}h_{Rast}(t)\\
\\
\delta z(t)=0.
\end{array}\label{eq: Ratsall case}
\end{equation}
Thus, Eqs. (\ref{eq: traditional GTR}), (\ref{eq: f(R) case}) and
(\ref{eq: Ratsall case}) can, in principle, be used in order to discriminate
among the GTR, $f(R)$ theories of gravity and the Rastall theory
of gravity. On the other hand, at the present time, the sensitivity
of the current ground based GW interferometers is not sufficiently
high to determine if the total displacements of the test mass are
governed by Eqs. (\ref{eq: traditional GTR}), or if they are governed
by Eqs. (\ref{eq: f(R) case}) or by Eqs. (\ref{eq: Ratsall case}).
A network including various interferometers in addition to LIGO and
Virgo with different orientations is indeed required. In fact, one
hopes that future advancements in ground-based projects and space-based
projects will have a sufficiently high sensitivity to determine, with
absolute precision, the direction of the GW propagation and the motion
of the various involved test masses. In other words, in the nascent
GW astronomy we hope not only to obtain new, precious astrophysical
information, but we also hope to be able to discriminate between Eqs.
(\ref{eq: traditional GTR}), Eqs. (\ref{eq: f(R) case}) and Eqs.
(\ref{eq: Ratsall case}).

\section{Conclusion remarks}

The era of the GW astronomy which recently started with the events
GW150914 \cite{key-1}, GW151226 \cite{key-2}, GW170104 \cite{key-3},
GW170814 \cite{key-48} and GW170817 \cite{key-49} is considered
a cornerstone for science and for gravitational physics in particular.
On one hand, GW astronomy permits to obtain new fundamental astrophysical
information from the Universe. On the other hand, it could ultimately
discriminate among the GTR and alternative gravitational theories.
At the present time, despite the cited events have put strong constrains
on the GTR, alternative gravitational theories are still viable. In
this paper we focused on the Rastall theory of gravity, a particular
extended theory of gravity which recently obtained an increasing interest
in the literature. Following a profound analogy between GWs in $f(R)$
theories and GWs in the Rastall theory, we linearized the Rastall
field equations and found the corresponding GWs. After that, the motion
of the test masses due to GWs in this theory has been also studied.
This could help, in principle, to discriminate between the GTR, $f(R)$
theories and the Rastall theory of gravity. In fact, the main result
of this paper is the system of equations (\ref{eq: Ratsall case})
which governs the total displacements of a test mass due to a GW in
Rastall theory. Despite the sensitivity of the current ground based
GW interferometers is not sufficiently high to determine if the total
displacements of the test mass are governed by such equations or by
Eqs. (\ref{eq: traditional GTR}), which are the traditional equations
governing the total displacements of the test mass due to a GW in
the standard GTR, or by Eqs. (\ref{eq: f(R) case}), which are the
traditional equations governing the total displacements of the test
mass due to a GW in $f(R)$ theories, we hope in future advancements
in ground-based projects and space-based projects.

As a final remark we start to discuss an intriguing issue, which could
be, in principle, developed in the future. Following the analogy between
$f(R)$ theories and the Rastall theory of gravity, in this paper
we have shown that also the Rastall theory admits a third GW polarization
like the case of $f(R)$ theories of gravity. The difference is that
in $f(R)$ theories such a third mode is massive while in the Rastall
theory it is massless. We have also seen that in both of the cases
the existence of the third polarization is due to the presence of
additional curvature terms in the gravitational action. Remarkably,
we know that, in general, all the extended theories of gravity should
have additional polarizations with respect to the standard $+$ and
$\times$ polarizations of the GTR, see for example \cite{key-54}.
Thus, one can ask which is the physical meaning of those additional
polarizations. Is there any general principle behind the existence
of such additional polarizations in extended gravity? On one hand,
we think that, in some way, the solution could be in the vacuum re-definition.
It could be necessary to give up the notion of ``perfect vacuum''
and this could enable a plurality of gauge conditions which must be
compatible each other. On the other hand, if one considers the particular
case of the Rastall theory one can see this theory as a relativistic
rewriting of Newtonian theory. Then, the two standard $+$ and $\times$
polarizations depends on the field structure of the GTR within the
Maxwellian approach, see \cite{key-33}, and the third mode should
arise from the Newtonian background. Developing these thoughts in
a rigorous way could be not banal and could require further profound
analysis, maybe searching analogies also with the Einstein-Cartan
theory {[}55 - 58{]}. In that case, the algebraic connection between
the Rastall and Einstein-Cartan theories should be the \textquotedbl{}freedom
degree\textquotedbl{} respect to the GTR, and it should be physically
equivalent to \textquotedbl{}double face\textquotedbl{} vacuum. Then,
also the Einstein-Cartan theory might have a third polarization model.

\section{Acknowledgements }

Christian Corda has been supported financially by the Research Institute
for Astronomy and Astrophysics of Maragha (RIAAM).

\end{document}